# Exploration of Various Fractional Order Derivatives in Parkinson's Disease Dysgraphia Analysis

Jan Mucha[1], Zoltan Galaz[1], Jiri Mekyska[1], Marcos Faundez-Zanuy[2], Vojtech Zvoncak[1], Zdenek Smekal[1], Lubos Brabenec[3], and Irena Rektorova[3,4]

[1] Department of Telecommunications, Faculty of Electrical Engineering and Communication, Brno University of Technology, Brno, Czech Republic
mucha@vut.cz

[2] Escola Superior Politecnica, Tecnocampus, Mataro, Barcelona, Spain

[3] Applied Neuroscience Research Group, Central European Institute of Technology – CEITEC, Masaryk University, Brno, Czech Republic

[4] First Department of Neurology, Faculty of Medicine and St. Anne's University Hospital, Masaryk University, Brno, Czech Republic
irena.rektorova@fnusa.cz

**Abstract.** Parkinson's disease (PD) is a common neurodegenerative disorder with a prevalence rate estimated to 2.0% for people aged over 65 years. Cardinal motor symptoms of PD such as rigidity and bradykinesia affect the muscles involved in the handwriting process resulting in handwriting abnormalities called PD dysgraphia. Nowadays, online handwritten signal (signal with temporal information) acquired by the digitizing tablets is the most advanced approach of graphomotor difficulties analysis. Although the basic kinematic features were proved to effectively quantify the symptoms of PD dysgraphia, a recent research identified that the theory of fractional calculus can be used to improve the graphomotor difficulties analysis. Therefore, in this study, we follow up on our previous research, and we aim to explore the utilization of various approaches of fractional order derivative (FD) in the analysis of PD dysgraphia. For this purpose, we used the repetitive loops task from the Parkinson's disease handwriting database (PaHaW). Handwritten signals were parametrized by the kinematic features employing three FD approximations: Grünwald-Letnikov's, Riemann-Liouville's, and Caputo's. Results of the correlation analysis revealed a significant relationship between the clinical state and the handwriting features based on the velocity. The extracted features by Caputo's FD approximation outperformed the rest of the analyzed FD approaches. This was also confirmed by the results of the classification analysis, where the best model

This work was supported by grant no. NU20-04-00294 (Diagnostics of Lewy body diseases in prodromal stage based on multimodal data analysis) of the Czech Ministry of Health and by Spanish grant of the Ministerio de Ciencia e Innovación no. PID2020-113242RB-I00 and by EU grant Next Generation EU (project no. LX22NPO5107 (MEYS)).

trained by Caputo's handwriting features resulted in a balanced accuracy of 79.73% with a sensitivity of 83.78% and a specificity of 75.68%.

## 1 Introduction

Fractional calculus (FC) is a name of the theory of integrals and derivatives of an arbitrary order [28]. It has been developed simultaneously with the well-known differential calculus [16] and its principles have been successfully used in modern engineering and science in general [18,32,37]. The advances of FC have been employed in the modeling of different diseases as well, like the human immunodeficiency virus (HIV) [2] or malaria [27]. In addition, the FC has been widely utilized in several computer vision disciplines such as the super-resolution, motion estimation, image restoration or image segmentation [34]. Furthermore, in our recent research we developed new handwriting features extraction techniques based on the application of the fractional order derivatives (FD) [11,21–25].

Parkinson's disease (PD) is a chronic idiopathic disorder, with the prevalence rate estimated to be approximately 2.0% for people aged over 65 years [12]. It is characterized by the progressive loss of dopaminergic neurons in the *substancia nigra pars compacta* [6,13], which is a major cause of the symptoms linked with the PD. Primary PD motor symptoms are tremor at rest, muscular rigidity, progressive bradykinesia, and postural instability [3,14]. One of the essential motor symptoms of PD is PD dysgraphia [17,36]. Additionally, a variety of non-motor symptoms such as cognitive impairment, sleep disturbances, depression, etc. may arise.

PD dysgraphia includes a spectrum of neuromuscular difficulties like motor-memory dysfunction, motor feedback difficulties, graphomotor production deficits and others [17,31]. These disabilities leads to a variety of handwriting difficulties manifesting as dysfluent, shaky, slow, and less readable handwriting. The most commonly observed handwriting abnormality in PD patients is micrographia. Micrographia represents the progressive decrease of letter's amplitude or width [20]. Some PD patients never develop micrographia, but they still exhibit other handwriting difficulties. Accordingly, the consequences of PD dysgraphia significantly affect a person's quality of life. Starting with slow and less legible handwriting and often progressing to lower self-esteem, poor emotional well-being, problematic communication and social interaction, and many others. Nowadays, the most advanced approaches of the PD manifestations quantification contained in the handwriting are based on digitizing tablets [9,21,35]. These devices can acquire x and y trajectories along with temporal information, therefore the temporal, kinematic, or dynamic characteristics can be processed together with the spatial features. Handwritten signal acquired by the digitizing tablet is called online handwriting.

In the past decades, researchers have been exploring the effect of several handwriting/drawing tasks in PD dysgraphia analysis, including the simplest ones (loops, circles, lines, Archimedean spiral) together with more complex ones (words, sentences, drawings, etc.) [7–9,21–23,26]. Drotar et al. [7–9] reported classification accuracy up to 89% using a combination of kinematic, pressure, energy or empirical mode decomposition features. The diagnosis of PD with accuracy of 71.95% based on the kinematic and entropy features extracted from the sentence task was reported by Impedovo et al. [15]. Taleb et al. [35] reported up to 94% accuracy of PD severity prediction using kinematic and pressure features in combination with adaptive synthetic sampling approach (ADASYN) for model training. Rios-Urrego et al. [30] achieved classification accuracy of 83.3% using the kinematic, geometric, spectral and nonlinear dynamic features. New kinematic features utilizing the discrete time wavelet transform, the fast Fourier transform and a Butter/adaptive filter introduced by Aouraghe et al. [1] resulted in classification accuracy of 92.2%.

Finally, in our recent works [21–23,25] we introduced and evaluated a new advanced approach of PD dysgraphia analysis employing the FD as a substitution of the conventional differential derivative during the basic kinematic feature extraction. Newly designed handwriting features achieved classification accuracy up to 90%, using the Grünwald-Letnikov approach only. In addition to PD dysgraphia analysis, we explored the FD-based handwriting features in analysis of graphomotor difficulties in school-aged children, where we examined three different FD approaches [24]. The results suggests that the employment of various FD approximations brings major differences in kinematic handwriting features. Therefore, as a next logical step, this study aims to:

1. extend our previous research in PD dysgraphia analysis by the utilization of various FD approaches,
2. explore the differences of various FD approaches in the analysis of PD dysgraphia,
3. compare the power of the FD-based handwriting features extracted by several FD approximations to distinguish between the PD patients and healthy controls (HC).

## 2 Materials and Methods

### 2.1 Dataset

For the purpose of this study, we used the Parkinson's disease handwriting database (PaHaW) [7]. The database consists of several handwriting or drawing tasks acquired in 37 PD patients and 38 healthy controls (HC). The participants were enrolled at the First Department of Neurology, St. Anne's University Hospital in Brno, Czech Republic. All participants reported Czech language as their native language and they were right-handed. The patients completed their tasks approximately 1 h after their regular dopaminergic medication (L-dopa). All participants signed an informed consent form approved by the local ethics committee. Demographic and clinical data of the participants involved in this study can be found in Table 1. For the purpose of this study, we selected the repetitive loop handwriting task. This task is missing for several participants of the PaHaW dataset, therefore, we processed 31 PD patients and 37 HC only.

**Table 1.** Demographic and clinical data of the participants.

| Gender | N | Age [y] | PD dur [y] | UPDRS V | LED [mg/day] |
|---|---|---|---|---|---|
| Parkinson's disease patients | | | | | |
| Females | 15 | 70.2 ± 8.4 | 7.9 ± 3.9 | 1.9 ± 0.4 | 1129.7 ± 572.9 |
| Males | 16 | 65.9 ± 13.1 | 7.0 ± 3.9 | 2.4 ± 0.9 | 1805.7 ± 743.3 |
| All | 31 | 68.0 ± 11.1 | 7.4 ± 3.9 | 2.2 ± 0.8 | 1478.6 ± 739.8 |
| Healthy controls | | | | | |
| Females | 17 | 61.6 ± 10.2 | – | – | – |
| Males | 20 | 63.3 ± 12.5 | – | – | – |
| All | 37 | 62.9 ± 11.5 | – | – | – |

N – number of subjects; y – years; PD dur – PD duration; UPDRS V – Unified Parkinson's disease rating scale, part V: Modified Hoehn & Yahr staging score [10]; LED – L-dopa equivalent daily dose.

## 2.2 Data Acquisition

The PaHaW database [7] consists of nine handwriting tasks. For the purpose of this study we selected the repetitive loop task only. An example of the repetitive loop task for a PD patient and a HC can be seen in Fig. 1. During the acquisition of the handwriting tasks, the participants were rested and seated in a comfortable position with a possibility to look at a pre-filled template. In case of some mistakes, they were allowed to repeat the task. A digitizing tablet (Wacom Intuos 4M) was overlaid with an empty paper and the participants wrote on that using the Wacom Inking pen. Online handwriting signals were recorded with $f_s$ = 150 Hz sampling rate, and the following time sequences were acquired: x and y coordinates ($x[t]$, $y[t]$); time-stamp ($t$); on-surface and in-air movement status ($b[t]$); pressure ($p[t]$); azimuth ($az[t]$); and tilt (also called altitude; $al[t]$).

## 2.3 Fractional Order Derivative

The main subject of this study is the exploration of the various FD approximations as a substitution of the conventional differential derivatives in the handwriting feature extraction process. We utilized three different FD approximations, namely: Grünwald-Letnikov (GL), Riemann-Liouville (RL), and Caputo (C), implemented by Valério Duarte in Matlab [38–40].

First approach employed in this study was developed by Grünwald and Letnikov. A direct definition of the derivation of the function $y(t)$ by the order $\alpha$ – $D^\alpha y(t)$ [28] is based on the finite differences of an equidistant grid in [0,$\tau$], assuming that the function $y(t)$ satisfies certain smoothness conditions in every finite interval $(0, t), t \leq T$, where T denotes the period. Choosing the grid

$$0 = \tau_0 < \tau_1 < \ldots < \tau_{n+1} = t = (n+1)h, \tag{1}$$

with

$$\tau_{k+1} - \tau_k = h, \tag{2}$$

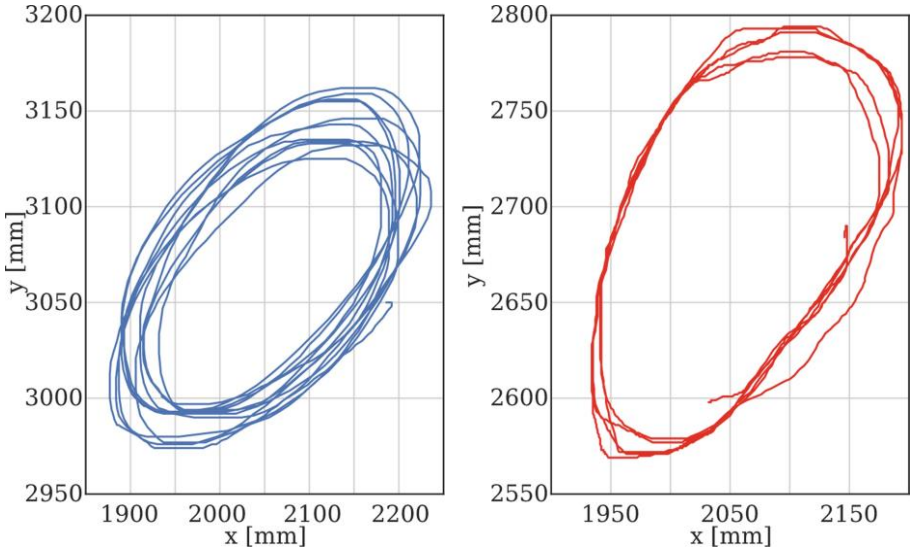

**Fig. 1.** Example of the repetitive loop task for a HC (left) and a PD patient (right).

and using the notation of finite differences

$$\frac{1}{h^\alpha}\Delta_h^\alpha y(t) = \frac{1}{h^\alpha}\left[y(\tau_{n+1}) - \sum_{v=1}^{n+1} c_v^\alpha y(\tau_{n+1-v})\right], \quad (3)$$

where

$$c_v^\alpha = (-1)^{v-1}\binom{\alpha}{v}. \quad (4)$$

The Grünwald–Letnikov definition from 1867 is defined as

$$^{GL}D^\alpha y(t) = \lim_{h \to 0} \frac{1}{h^\alpha}\Delta_h^\alpha y(t), \quad (5)$$

where $^{GL}D^\alpha y(t)$ denotes the Grünwald-Letnikov derivatives of order $\alpha$ of the function $y(t)$, and $h$ represents the sampling lattice.

Second approach used in this study has been given by Riemann-Liouville. The left-inverse interpretation of $D^\alpha y(t)$ by Riemann-Liouville [18,28] from 1869 is defined as

$$^{RL}D^\alpha y(t) = \frac{1}{\Gamma(n-\alpha)}\frac{d^n}{dt^n}\int_0^t (t-\tau)^{n-\alpha-1}y(t)\,dt, \quad (6)$$

where $^{RL}D^\alpha y(t)$ denotes the Riemann-Liouville derivatives of order $\alpha$ of the function $y(t)$, $\Gamma$ is the gamma function and $n-1 < \alpha \leq n, n \in \mathbf{N}, t > 0$.

Third and last FD approach involved in this study was developed by M. Caputo [4]. In contrast to the previous ones, the improvement hereabouts lies in

the unnecessity to define the initial FD condition [18,28]. The Caputo's definition from 1967 is

$$^{C}D^{\alpha}y(t) = \frac{1}{\Gamma(n-\alpha)} \int_0^t (t-\tau)^{n-\alpha-1} y^n(t) \, dt, \quad (7)$$

where $^{C}D^{\alpha}y(t)$ denotes the Caputo derivatives of order $\alpha$ of the function $y(t)$, $\Gamma$ is the gamma function and $n-1 < \alpha \leq n, n \in \mathbf{N}, t > 0$.

### 2.4 Feature Extraction

Considering the nature of the selected task, on-surface handwriting features were extracted only. Since we did employ three FD approaches in the feature extraction process, three sets of the handwriting features were created. Digitizing tablet rarely omits 3–4 samples during the acquisition, therefore the in-signal outliers removal was performed (outliers were considered as elements more than three scaled median absolute deviations from the median). If not pre-processed, the differentiation of this gap would leave significant peaks in the output handwriting feature. All handwriting features were computed for $\alpha$ in the range of 0.1–1.0 (with the step of 0.1), where $\alpha$ = 1.0 is equal to the full derivation. Furthermore, the statistical properties of all extracted handwriting features were described by the mean and the relative standard deviation (relstd). To sum up, each feature set consists of 180 computed kinematic features.

### 2.5 Statistical Analysis and Machine Learning

Firstly, the normality test of the handwriting features using the Shapiro-Wilk test was performed [33]. Since most of the features were found to come from normal distribution, we did not apply any normalization on a feature basis. To control for the effect of confounding factors (also known as covariates), we controlled for the effect of age and gender of the subjects.

Next, Spearman's ($\rho$) and Pearson's ($r$) correlation coefficient with the significance level of 0.05 were computed to assess the strength of the monotonous and linear relationship between the handwriting features and the subject's clinical status (PD/HC). Finally, to control for the issue of multiple comparisons, p-values were adjusted using the False Discovery Rate (FDR) method.

Consequently, binary classification models were built in order to distinguish between the PD patients and HC utilizing the extracted handwriting features. An ensemble extreme gradient boosting algorithm known as XGBoost [5] (with 100 estimators) was used for this purpose. The XGBoost algorithm was selected due to its ability to find complex interactions among features as well as the possibility of ranking their importance and its robustness to outliers. Hyper-parameter space optimization (1000 iteration) by the randomized search strategy (stratified 5-fold cross-validation with 10 repetitions) was performed to optimize balanced accuracy. The set of hyper-parameters that were optimized can be found in the following table (Table 2).

Table 2. Hyper-parameters set.

| Hyper-parameter | Values |
|---|---|
| Learning rate | [0.001, 0.01, 0.1, 0.2, 0.3] |
| Gamma | [0, 0.05, 0.10, 0.15, 0.20, 0.25, 0.5] |
| Maximum tree depth | [6, 8, 10, 12, 15] |
| Subsample ratio | [0.5, 0.6, 0.7, 0.8, 0.9, 1.0] |
| Columns subsample ratio at each level | [0.4, 0.5, 0.6, 0.7, 0.8, 0.9, 1.0] |
| Columns subsample ratio for each tree | [0.4, 0.5, 0.6, 0.7, 0.8, 0.9, 1.0] |
| Balance between positive and negative weights | [1, 2, 3, 4] |
| Minimum weights required in a child node | [0.5, 1.0, 3.0, 5.0, 7.0, 10.0] |

The classification performance was evaluated by the following classification metrics: Matthew's correlation coefficient [19] (MCC), balanced accuracy (BACC), sensitivity (SEN) also known as recall (REC), specificity (SPE), precision (PRE) and F1 score (F1). These metrics are defined as follows:

$$\text{MCC} = \frac{TP \times TN + FP \times FN}{\sqrt{N}}, \tag{8}$$

$$\text{BACC} = \frac{1}{2}\left(\frac{TP}{TP+FN} + \frac{TN}{TN+FP}\right), \tag{9}$$

$$\text{SPE} = \frac{TN}{TN+FP}, \tag{10}$$

$$\text{PRE} = \frac{TP}{TP+FP}, \tag{11}$$

$$\text{REC} = \frac{TP}{TP+FN}, \tag{12}$$

$$\text{F1} = 2\frac{PRE \times REC}{PRE+REC} \tag{13}$$

where $N = (TP + FP) \times (TP + FN) \times (TN + FP) \times (TN + FN)$, $TP$ (true positive) and $FP$ (false positive) represent the number of correctly identified PD patient and the number of subjects incorrectly identified as PD patient, respectively. Similarly, $TN$ (true negative) and $FN$ (false negative) represent the number of correctly identified HC and the number of subjects with PD incorrectly identified as being healthy.

For a better illustration, the overview of the performed analysis from the handwriting task selection to the evaluation of the results can be found in Fig. 2.

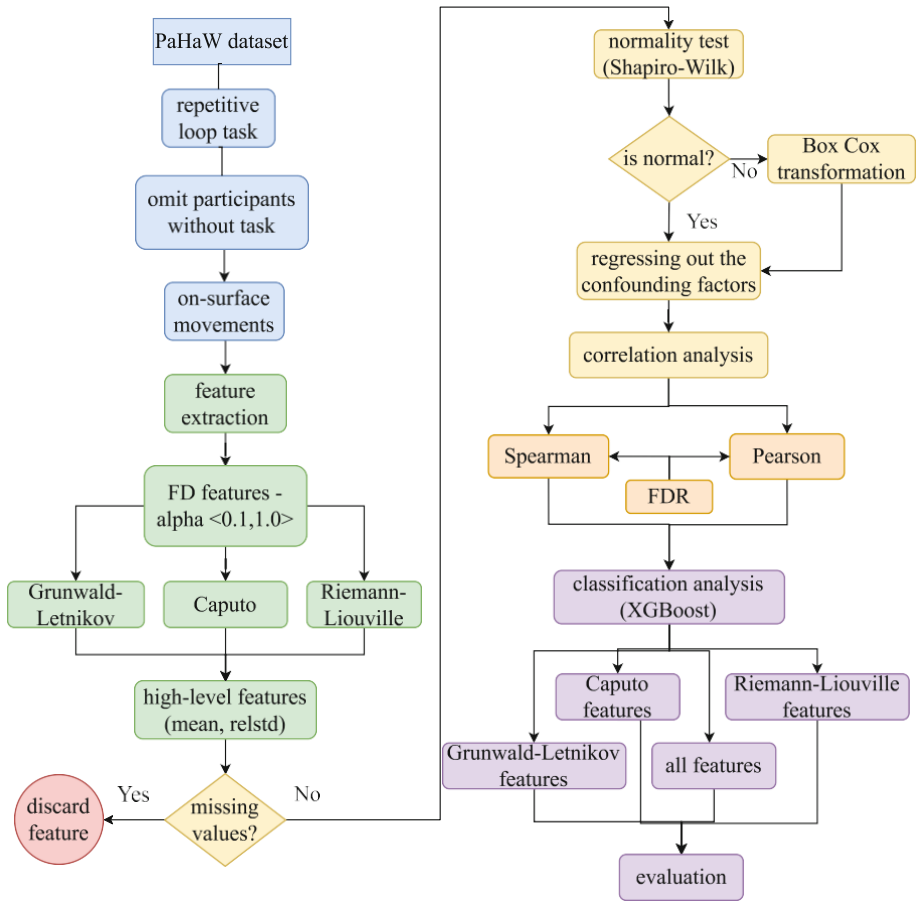

**Fig. 2.** Flow overview of the performed experiments.

## 3 Results

The results of the correlation analysis can be seen in Table 3, where the top 5 features per FD approximation according to the p-values of Spearman's correlation are shown. The most significant correlation (after the FDR adjustment) with the clinical state (PD/HC) of the participants was identified in the features extracted by the Caputo's FD approach. Nevertheless, all FD approaches provided the handwriting features that pass the selected significance level ($p < 0.05$), while features extracted by Caputo's and Riemann-Liouville's achieved the p-values very close to 0. Most of the top selected handwriting features are based on horizontal velocity, and all of them have $\alpha$ different from 1, which confirms the positive impact of the FD in PD dysgraphia analysis.

The results of the classification analysis are summarized in Table 4. In total, 4 models were trained: one model per each FD approach and one model com-

**Table 3.** Results of the correlation analysis between the subjects' clinical status (PD/HC) and the computed handwriting features ranked by the adjusted p-value (and the correlation coefficient) of Spearman's correlation.

| Feature name | $\rho$ | $p_s$ | $p_s^*$ | $r$ | $p_p$ | $p_p^*$ |
|---|---|---|---|---|---|---|
| Caputo | | | | | | |
| relstd horizontal velocity-$\alpha$ = 0.6 | −0.5408 | 0.0001 | 0.0001 | −0.5456 | 0.0001 | 0.0001 |
| relstd horizontal velocity-$\alpha$ = 0.5 | −0.5122 | 0.0001 | 0.0001 | −0.5204 | 0.0001 | 0.0001 |
| relstd horizontal velocity-$\alpha$ = 0.4 | −0.4912 | 0.0001 | 0.0001 | −0.5024 | 0.0001 | 0.0001 |
| mean horizontal velocity-$\alpha$ = 0.3 | 0.4791 | 0.0001 | 0.0001 | 0.4049 | 0.0006 | 0.0051 |
| mean horizontal velocity-$\alpha$ = 0.4 | 0.4716 | 0.0001 | 0.0001 | 0.4240 | 0.0003 | 0.0036 |
| Grünwald-Letnikov | | | | | | |
| relstd horizontal velocity-$\alpha$ = 0.8 | −0.4475 | 0.0001 | 0.0180 | −0.4332 | 0.0002 | 0.0240 |
| relstd horizontal velocity-$\alpha$ = 0.9 | −0.4310 | 0.0002 | 0.0180 | −0.4184 | 0.0004 | 0.0240 |
| relstd horizontal velocity-$\alpha$ = 0.7 | −0.4220 | 0.0003 | 0.0180 | −0.4162 | 0.0004 | 0.0240 |
| relstd horizontal velocity-$\alpha$ = 0.6 | −0.3964 | 0.0008 | 0.0324 | −0.3682 | 0.0020 | 0.0720 |
| relstd vertical velocity-$\alpha$ = 0.9 | −0.3949 | 0.0009 | 0.0324 | −0.3801 | 0.0014 | 0.0630 |
| Riemann-Liouville | | | | | | |
| mean horizontal velocity-$\alpha$ = 0.2 | 0.4882 | 0.0001 | 0.0001 | 0.3869 | 0.0011 | 0.0060 |
| relstd horizontal velocity-$\alpha$ = 0.2 | −0.4716 | 0.0001 | 0.0001 | −0.4643 | 0.0001 | 0.0013 |
| mean horizontal velocity-$\alpha$ = 0.3 | 0.4716 | 0.0001 | 0.0001 | 0.4240 | 0.0003 | 0.0022 |
| relstd vertical velocity-$\alpha$ = 0.2 | −0.4686 | 0.0001 | 0.0008 | −0.4654 | 0.0001 | 0.0013 |
| relstd vertical velocity-$\alpha$ = 0.3 | −0.4475 | 0.0001 | 0.0008 | −0.4483 | 0.0001 | 0.0013 |

$\rho$ – Spearman's correlation coefficient; $p_s$ – p-value of Spearman's correlation; $p_s^*$ – adjusted p-value of Spearman's correlation; $r$ – Pearson's correlation coefficient; $p_p$ – p-value of Pearson's correlation; $p_p^*$ – adjusted p-value of Pearson's correlation; relstd – relative standard deviation; h. – horizontal; v. – vertical.

bining all the features. The best classification performance was achieved by the Caputo's FD approach with BACC = 0.7973, SEN = 0.8378, SPE = 0.7568, PRE = 0.7750 and F1 = 0.8052. However, the highest SEN and SPE were achieved by the Riemann-Liouville approach (SPE = 0.8378, PRE = 0.8065).

Next, in Fig. 3 the comparison of the horizontal velocity function for $\alpha$ = 0.6 across all of the utilized FD approximations is visualized. The handwriting features were extracted from the performance of the PD patient with high PD severity. And finally, an example of the dependency of the mean of horizontal velocity on the FD order $\alpha$ for all three FD approaches is shown in Fig. 4.

## 4  Discussion

The main goal of this study is to explore various FD approximations and their differences in the analysis of the PD dysgraphia by online handwriting. For better illustration and more understanding of the differences as well as the common

**Table 4.** Results of the classification analysis.

| FD approach | MCC | BACC | SEN | SPE | PRE | F1 |
|---|---|---|---|---|---|---|
| C | **0.5966** | **0.7973** | **0.8378** | 0.7568 | 0.7750 | **0.8052** |
| RL | 0.5204 | 0.7568 | 0.6757 | **0.8378** | **0.8065** | 0.7353 |
| GL | 0.4867 | 0.7432 | 0.7297 | 0.7568 | 0.7500 | 0.7397 |
| ALL | 0.5135 | 0.7568 | 0.7568 | 0.7568 | 0.7568 | 0.7568 |

MCC – Matthew's correlation coefficient; BACC – balanced accuracy; SEN – sensitivity; SPE – specificity; PRE – precision; F1 – F1 score; GL – Grünwald-Letnikov; C – Caputo; RL – Riemann-Liouville; ALL (combination of all feature-types, i. e. 540 features).

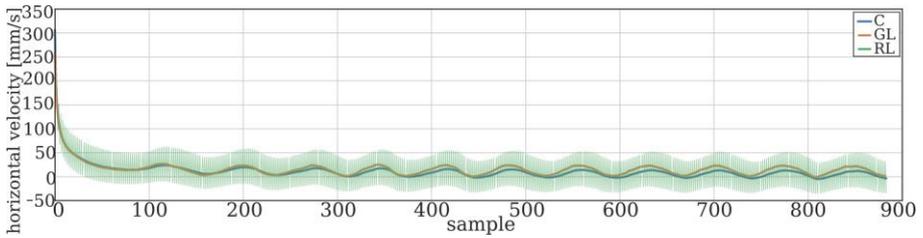

**Fig. 3.** Comparison of the horizontal velocity function ($\alpha = 0.6$) across all of the FD approximations (PD patient; C – Caputo; GL – Grünwald-Letnikov; RL – Riemann-Liouville).

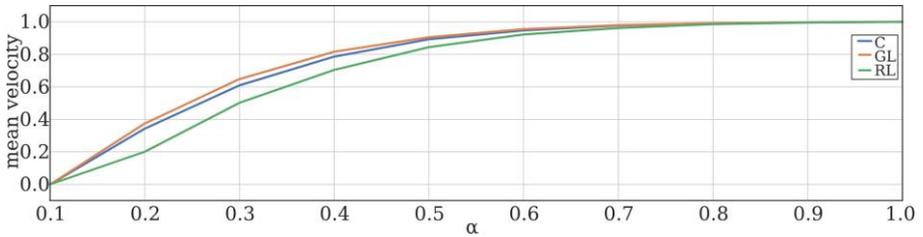

**Fig. 4.** Mean of horizontal velocity depending on FD order $\alpha$ (PD patient; C – Caputo; GL – Grünwald-Letnikov; RL – Riemann-Liouville).

characteristics, the comparison of the identical handwriting feature extracted for all three FD approaches can be found in Fig. 3. The feature is extracted from the handwritten product of a PD patient and the feature represents the horizontal velocity for $\alpha = 0.6$. The velocity function extracted by the Riemann-Liouville's approximation dominates by its oscillatory nature in comparison to the other two approaches. Nevertheless the envelope of Riemann-Liouville's approach follows the local maximums and minimums of the functions computed by the Caputo's and Grünwald-Letnikov's approximation. A minor shift of the velocity function can be noticed between the Caputo's and Grünwald-Letnikov's approaches. This is due to the nature of the Caputo's FD approach, which differentiates input data before the convolution operation, so the temporal memory is applied to the velocity afterwards. Regarding the visualization in Fig. 3, we can confirm the differences in the same handwriting feature extracted by various FD approximations. Additionally, the dependency comparison of the mean of horizontal velocity on the order $\alpha$ is provided in Fig. 4. The oscillatory behaviour of the Riemann-Liouville's function results in the wider gap from the Caputo's and Grünwald-Letnikov's functions. Nevertheless, all three FD approaches converge to the same point as the order $\alpha$ is closer to 1.0. This behaviour is expected, because the full derivation has to be the same for all approaches.

Regarding the results of the correlation analysis, the most significantly correlated handwriting features (after the FDR adjustment) were extracted by the Caputo's FD. This observation is in line with our previous results [24], where we analysed the same three FD approaches in assessment of the graphomotor difficulties in school-aged children. The performance of the handwriting features extracted by the Riemann-Liouville's approach is almost as good as the Caputo's features. The Grünwald-Letnikov's handwriting features achieved weaker relationship, however the features are still below selected level of significance ($p < 0.05$). Most significantly correlated handwriting features are related to the horizontal velocity. In general, PD dysgraphia is linked with the reduced velocity, which could occur even more often than micrographia [15,29,31]. This strong relationship is reasonable due to the cardinal symptoms of PD, such as bradykinesia or rigidity, which have a significant impact on fine motor skills, including handwriting/drawing. Moreover, some studies suggest that the horizontal version of micrographia is even more common than the vertical version [36]. The values of the correlation coefficients for handwriting features described by the mean are positive, which means that the performance of the participant is worse with the higher values of the horizontal velocity. This can be confusing because just the opposite effect may be expected. However, this may be specific for the repetitive loop task, where the velocity for the healthy writer is more constant. On the other hand, the writer with PD dysgraphia performs the loop more jerkily, which leads to higher velocity with more variability. This is confirmed by the fact that the features described by the relative standard deviation are negative, which means that the handwriting performance is better with the lower variability of the horizontal velocity.

Based on the results of the classification analysis, the best classification performance was obtained by the handwriting features computed by Caputo's FD. The resulting balanced accuracy was 79.73% with SEN = 83.78% and SPE = 75.68%. In our similar study [21] we achieved classification accuracy of 80.60% with SEN = 79.4% and SPE = 80.56% using all of the handwriting tasks from the PaHaW database, but only the Grünwald-Letnikov FD was employed. In comparison to this study, we can conclude that the exploration of the various FD approaches improved the classification analysis, considering that we achieved almost the same performance only by one handwriting task and using the on-surface kinematic features only. The balanced accuracy of the Riemann-Liouville and Grünwald-Letnikov FD is approximately 5% lower while the sensitivity is lower up to 15% in comparison to the Caputo's FD. Considering the reported results, we can conclude that the Caputo's approach is the most suitable FD approximation of the kinematic analysis of the PD dysgraphia by online handwriting.

## 5 Conclusion

To the best of our knowledge, this is one of the first studies performing an investigation of the various FD approaches in the computerized analysis of the PD dysgraphia by online handwriting. For that reason, the outcomes should be considered as being rather exploratory and pilot in nature. Based on the reported results, Caputo's FD approximation outperformed the rest of the analysed FD approaches in all experiments. The correlation analysis resulted in the significant relationship between the clinical state and the handwriting features based on the velocity, which is in line with our previous findings. Additionally, the best classification model achieved the balanced accuracy of 79.73% with SEN = 83.78% and SPE = 75.68%, which is a comparable result to our previous studies.

This study has several limitations and possible parts, that could be further improved. The processed dataset is relatively small in terms of the statistical validity of the achieved results. Next, the $\alpha$ order should be explored more sensitively (e.g. with a step of 0.01 or even less) in order to identify the optimal range for PD dysgraphia analysis. Additionally, other feature types, such as temporal, spatial, and dynamic, should be included in future comparisons. Moreover, the comparison of the various FD-based features with the conventionally used handwriting features should be performed. Besides, all handwriting tasks included in the PaHaW database have to be investigated by the various FD approaches. And finally, various machine learning models should be trained and compared in future studies.